\documentclass[aps,preprint,showkeys,a4paper,tightenlines,superscriptaddress,12pt]{revtex4}

\usepackage{eurosym}
\usepackage{amsfonts}
\usepackage{amsmath}
\usepackage{amssymb}
\usepackage{graphicx}
\usepackage{subfig}
\usepackage{caption}
\usepackage{setspace}

\providecommand{\U}[1]{\protect\rule{.1in}{.1in}}

\begin{document}

\section*{SUBMITTED TO "MECHANICS OF MATERIALS" ON 4th OCTOBER 2020}

\title{Rate-dependent adhesion of viscoelastic contacts. Part \textrm{I}:
contact area and contact line velocity within model multi-asperity contacts with rubber.}
\author{G. Violano}
\affiliation{Department of Mechanics, Mathematics and Management, Polytechnic
University of Bari, Via E. Orabona, 4, 70125, Bari, Italy}
\author{A. Chateauminois}
\affiliation{Soft Matter Science and Engineering Laboratory (SIMM), PSL Research
University, UPMC Univ Paris 06, Sorbonne Universités, ESPCI Paris, CNRS, 10 rue
Vauquelin, 75231 Paris cedex 05, France}
\author{L. Afferrante}
\email{guido.violano@poliba.it}
\affiliation{Department of Mechanics, Mathematics and Management, Polytechnic
University of Bari, Via E. Orabona, 4, 70125, Bari, Italy}

\begin{abstract}
In this work, we investigate dissipative effects involved during the detachment of a smooth spherical glass probe from a viscoelastic silicone substrate patterned with micro-asperities. As a baseline, the pull-off of a single asperity, millimeter-sized contact between a glass lens and a smooth poly(dimethylsiloxane) (PDMS) rubber is first investigated as a function of the imposed detachment velocity. From a measurement of the contact radius $a(t)$ and normal load during unloading, the dependence of the strain energy relase rate $G$ on the velocity of the contact line $v_c=da/dt$ is determined under the assumption that viscoelastic dissipation is localized at the edge of the contact. These data are incoproated into Muller's model (V.M. Muller \textit{J Adh Sci Tech} (1999) \textbf{13} 999-1016) in order to predict the time-dependence of the contact size. Similar pull-off experiments are carried out with the same PDMS substrate patterned with spherical micro-asperities with a prescribed height distribution. From \textit{in situ} optical measurements of the micro-contacts, scaling laws are identified for the contact radius $a$ and the contact line velocity $v_c$. On the basis of the observed similarity between macro and microscale contacts, a numerical solution is developed to predict the reduction of the contact radius during unloading.
\end{abstract}


\maketitle

\section{Introduction}

Adhesion is of paramount importance in the contact mechanics of micro and
nano systems \cite{vakis2018} as, at the molecular scale, adhesive
interactions between atoms are `strong' compared to the usual forces acting
between bodies \cite{kendall}. However, adhesion is seldom observed at the
macroscopic scale due to surface roughness, which reduces the area of real
contact. Nevertheless, when dealing with very soft matter, strong adhesion
may be still detected even in presence of surface roughness \cite{tiwary2017}%
.

Soft matter adhesion finds applications in several fields, e.g. design of
pressure-sensitive-adhesives (PSA) \cite{PSA}, soft robots \cite{SOFTrobots}
and new technologies inspired by biotribological systems \cite{Biotribology}.

In most of adhesion tests on soft compliant spheres \cite%
{tiwary2017,Lorenz2013}, the measured detachment force is generally greatly
in excess of that predicted by Johnson, Kendall \& Roberts (JKR) theory \cite%
{JKR} and the detachment process is observed to be dependent on the rate of
separation \cite{GreenJohn1981}.

The JKR theory applies for purely elastic spheres and under quasi-static
conditions. In experimental investigations, the pull-off process unlikely
obeys the quasi-static conditions and the effective work of adhesion $\Delta
\gamma _{\mathrm{eff}}$ depends on the velocity $v_{\mathrm{c}}$ of the
contact line during pull-off. Namely, $\Delta \gamma _{\mathrm{eff}}$ may be
strongly increased with respect to the quasi-static value $\Delta \gamma
_{0} $ as a result of viscous dissipation, where $\Delta \gamma _{0}$
follows the well-known Dupr\'e's equation $\Delta \gamma _{\mathrm{0}}=\gamma
_{\mathrm{1}}+\gamma _{\mathrm{2}}-\gamma _{\mathrm{12}}$, being $\gamma _{%
\mathrm{1}}$, $\gamma _{\mathrm{2}}$ the adhesive energies of the two
contacting surfaces and $\gamma _{\mathrm{12}}$ the interaction term.

Gent \& Schultz (GS) \cite{GS1972} observed that viscous effects are
exclusively located close to the crack tip. Maugis \& Barquins (MB) \cite%
{MB1978} proposed a generalization of the JKR theory, showing that the
dependence of $\Delta \gamma _{\mathrm{eff}}$ on $v_{\mathrm{c}}$ can be
expressed in terms of a dissipation function\ $f(v_{\mathrm{c}},T)$ related
to the viscoelastic properties of the material and depending on the crack
tip velocity $v_{\mathrm{c}}$ and the temperature $T$. In particular, MB
showed that, for a given elastomer, the effective work of adhesion $\Delta
\gamma _{\mathrm{eff}}$ is a universal function of the crack tip velocity $%
v_{\mathrm{c}}$. Moreover, performing experimental tests on three different
geometries (spheres, punches and tapes (peeling)), MB found that the
dependence of $\Delta \gamma _{\mathrm{eff}}$ on $v_{\mathrm{c}}$ is not
affected by the geometry and loading system. In MB's solution, viscous
effects are assumed not involving bulk deformations as \textquotedblright 
\textit{gross displacements must be elastic for }$G$\textit{\ to be valid in
kinetic phenomena}\textquotedblright , being $G$ the energy release rate,
i.e. the amount of energy required to advance a fracture plane by a unit
area. Robbe-Valloire \& Barquins \cite{valloire1998} extended
MB studies performing adherence experiments between a rigid cylinder and an
elastomeric solid. They confirmed the existence of a master curve for $f(v_{%
\mathrm{c}},T)$. Specifically, their results "\textit{prove once again that
the master curve drawn and its variation... is a characteristic of the
propagation in mode I at the interface of our rubber-like material, when
viscoelastic losses are closely limited to the crack tip, so that G can be
calculated from the theory of linear elasticity.}"

More recently, Muller \cite{Muller1999} showed that the process of
detachment of viscoelastic spheres can be described by a first-order
differential equation, whose solution is based on the assumption originally
proposed in Ref. \cite{GS1972}. Alternative approaches taking into account
bulk deformations were proposed by the group of Barthel in Refs. \cite%
{Barthel2002,Barthel2003,Barthel2009}.

In this work, we present an experimental investigation of dissipative effects involved in the adhesion between a rough contact interface between a smooth spherical glass probe and a viscoelastic silicone substrate patterned with a prescribed height distribution of micrometer sized spherical asperities. Taking advantage from the fact that the size of these micro-asperities (radius of 100~$\mu$m) allows for an optical measurement of the space distribution of micro-contact areas, such patterned surfaces obtained from micro-milling techniques were previously successfully used to probe the elastic interactions between micro-asperity contacts~\cite{YASHIMA2015} or to investigate adhesive equilibrium of rough contact interfaces~\cite{ACITO2019}. Here, we focus on the effects of viscoelastic dissipation on the rate-dependence of of micro-contact sizes during unloading at a imposed velocity using JKR-type experiments. The investigation of adhesive behaviour at the level of micro-contact spots is inspired by the Roberts' statement (Ref. \cite{roberts79}): "\textit{The contact of a smooth centimerer-sized rubber sphere may be regarded as that of a giant single asperity...The ability to predict the adhesion forces on a large asperity is a step towards building up a model of a real surface of micron-sized asperities, which approximate to an array of minute hemispheres of different height and radius}".\\
Accordingly, pull-off experiments have first been conducted on smooth PDMS\ surfaces to investigate the adhesion behavior at the macroscopic scale. Specifically, under the assumption of viscous effects located only near the detachment front, we propose a very simple methodology to calculate the time-dependent radius of  the contacts during unloading by exploiting the Muller's approach \cite{Muller1999} which was already used to calculate the adhesion hysteresis occurring in loading-unloading tests performed on smooth viscoelastic spheres \cite{ViolAIAS2019}. Then, this approach is successfully extended at the micro-scale with no need to incorporate a size-dependence in the relationship ruling the dependence of the strain energy release rate on the velocity of the contact line.

\section{Detachment of viscoelastic spheres}

In order to detach a soft body from a rigid substrate, the force required to
create a new unit length of crack is $\left( G-\Delta \gamma _{\mathrm{0}%
}\right) $. If the energy release rate $G$ is larger than the adiabatic work
of adhesion $\Delta \gamma _{\mathrm{0}}$, the crack opens and the
detachment process advances.

Gent \& Schultz (GS) \cite{GS1972} found that%
\begin{equation}
G-\Delta \gamma _{\mathrm{0}}=\Delta \gamma _{\mathrm{0}}\cdot f(v_{\mathrm{c%
}},T)  \label{G1}
\end{equation}%
where $\Delta \gamma _{\mathrm{0}}\cdot f(v_{\mathrm{c}},T)$ is the drag due
to viscoelastic losses at the crack tip, being $v_{\mathrm{c}}=-da/dt$ the
velocity of the contact line. The above relation usually works for $v_{%
\mathrm{c}}$ ranging form $10^{-5}$ \textrm{cm/s}\ to $1$ \textrm{cm/s} \cite%
{Andrews73,Gent69,Kendall73,roberts79} and allows to predict the kinetics of
detachment (see Maugis \& Barquins \cite{MB1978}).

The function $f(v_{\mathrm{c}},T)$, which is found to be independent of the
geometry and loading system, can be described by the phenomenological
equation%
\begin{equation}
f(v_{\mathrm{c}})=k(a_{\mathrm{T}}v_{\mathrm{c}})^{n}\text{,}  \label{fvc}
\end{equation}%
where $k$ and $n$ are characteristic constants of the material and $a_{%
\mathrm{T}}$ is the William-Landel-Ferry (WLF)\ factor \cite{WLF1955}
accounting for the dependence of $f(v_{\mathrm{c}},T)$ on the temperature $T$%
. Eq. (\ref{fvc}) also accounts for the dependence of $G$ on the relaxed
elastic modulus $E$ (Ref. \cite{ramond1985}), whose frequency dependence
appears only at the crack tip \cite{charmet1998}.

Introducing eq. (\ref{fvc}) in (\ref{G1}), we obtain%
\begin{equation}
G=\Delta \gamma _{0}[1+c\cdot v_{\mathrm{c}}{}^{n}]\text{,}  \label{G2}
\end{equation}%
with $c=k\cdot a_{\mathrm{T}}{}^{n}$.

For a given elastomer, the values of $c$ and $n$ can be obtained by fitting
the experimental data relating $G$ and $v_{\mathrm{c}}$. As observed in Ref. \cite{Lorenz2013},
the exponent $n$ "\textit{is not a universal number, but takes different values depending on viscoelastic modulus}".

\subsection{Muller's model}

Muller \cite{Muller1999} proposed a two-parameters differential equation to
describe the detachment of a viscoelastic sphere of radius $R$ and Young
modulus $E^{\ast }$ from a rigid substrate%
\begin{equation}
\frac{d\bar{a}}{d\bar{\delta}}=\left[ \frac{\Delta \gamma _{\mathrm{0}}}{%
RE^{\ast }}\right] ^{1/3}\cdot \frac{1}{\beta }\left[ \bar{a}^{3}\left( 1-%
\frac{\bar{\delta}}{3\bar{a}^{2}}\right) ^{2}-\frac{4}{9}\right] ^{1/n}\text{%
,}  \label{Meq}
\end{equation}%
where $\bar{a}=a/\left[ 3R\left( \pi \Delta \gamma _{\mathrm{0}}/(6E^{\ast
}R)\right) ^{1/3}\right] $ and $\bar{\delta}=\delta /\left[ 3R\left( \pi
\Delta \gamma _{\mathrm{0}}/(6E^{\ast }R)\right) ^{2/3}\right] $ are the
dimensionless contact radius and penetration, respectively, and the
parameter $\beta $ is proportional to the driving velocity $V$%
\begin{equation}
\beta =\left( \frac{6}{\pi }\right) ^{1/3}\left( \frac{4}{9}c\right) ^{1/n}V%
\text{.}  \label{beta}
\end{equation}

Muller's model moves from two assumptions: i) viscous effects are located
exclusively near the crack tip; ii) detachment occurs under constant $V$.
The energy release rate $G$, which represents the effective work of adhesion 
$\Delta \gamma _{\mathrm{eff}}$ required to break the contact, can be
evaluated as%
\begin{equation}
G=\frac{\left( F_{\mathrm{H}}-F\right) ^{2}}{6\pi RF_{\mathrm{H}}}
\label{G3}
\end{equation}%
where $F_{\mathrm{H}}=4/3E^{\ast }a^{3}/R$ is the Hertzian load and $F$ is
the applied load. Eq. \ref{G3} is only valid under the assumption of viscous effects concentrated at the crack tip (see, for example, Ref. \cite{baekPDMS}).

\section{Experimental set-up}

JKR-like tests were carried out between an optical spherical glass lens and rubber substrates. The glass indenter, which is assumed
to be smooth, has a radius of curvature $R_{\mathrm{sphere}}$ of $103.7$ \textrm{mm}. Rubber substrates are made of  commercially available PolyDiMethylSiloxane (PDMS) silicones.
Samples were manufactured by cross-linking at $70$ $\mathrm{%
{{}^\circ}%
C}$ for $48$ hours a mixture of Sylgard $184$ and Sylgard $527$
silicones (Dow Chemicals), with a $0.35$\textrm{:}$0.65$ weight ratio. As detailed by Palchesko et al. \cite{Palchesko}, mixing these two silicone elastomers in different ratios allows to tune the elastic modulus in between a few kPa and 3~MPa. As compared to raw Sylgard $184$, the Young's modulus of the selected Sylgard $184$:Sylgard $527$ mixture ($E=$~0.83 MPa, see below) was decreased by a factor of about $3.6$, with the aim of enhancing the adhesion properties. In addition, crosslinking simultaneously these two different silicone products was expected to result in an increased concentration of network defects such as dangling chains. As detailed in Ref. \cite{Palchesko}, such imperfections are known to enhance the viscoelastic dissipation of silicone networks.\\

\begin{figure}[tbp]
\begin{center}
\includegraphics[width=15.0cm]{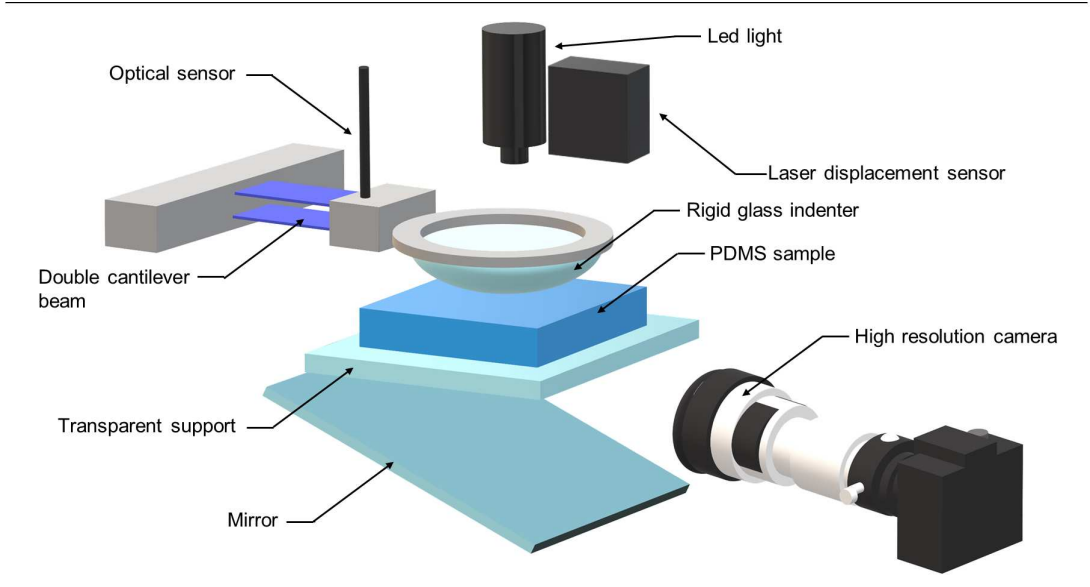}
\end{center}
\caption{The experimental setup of JKR adhesion tests.}
\label{Setup}
\end{figure}

Fig. \ref{Setup} shows a sketch of the experimental set-up. The spherical
indenter is fixed to a motorized vertical translation stage by means of a double
cantilever beam of known stiffness (290 \textrm{N m}$^{-1}$). The value of
the applied load is obtained from the deflection of the cantilever, as it is
measured using a high resolution optical sensor (Philtec D64-L). Due to the compliance of the double cantilever beam, the actual velocity of the lens can slightly different from the prescribed velocity. In order to account for this effect, a laser displacement sensor (Keyence LK-H057) is used to monitor the actual position of the lens. The difference between the prescribed and actual velocity of the indenter were found to be significant only for the macroscopic single asperity contact close to pull-off, when the greatest tensile normal forces are achieved.\\
The PDMS sample is fixed to two crossed motorized linear translation stages, which allow to change the relative position of the rubber sample with respect to the indenter. A LED light spot is installed to illuminate in transmission the contact area. Once illuminated, contact pictures are recorded through the transparent PDMS using a zoom objective (Leica APO Z16) and a high resolution CMOS camera (SVS Vistek Exo, $2048$ $\times $ $2048$ pixels$^{2}$, 8 bits).

\subsubsection{Experiments on smooth samples}

The contact radius vs. load data obtained by indentation experiments on smooth
PDMS\ samples were fitted according to the JKR theory \cite{JKR} in order to evaluate
the reduced elastic modulus ($E^{\ast }=0.83$ $\mathrm{MPa}$) and the
adhesion energy ($\Delta \gamma _{0}=0.037$ $\mathrm{J/m}^{2}$). During
loading, contact tests have been performed under fixed load conditions.
Specifically, the applied load is increased step by step and, once each load
step is reached, contact is maintained for a long time ($800$ \textrm{s}) to
ensure that adhesive equilibrium is reached (with viscoelastic effects
totally relaxed \cite{ACITO2019}).

Unloading tests are performed at imposed driving velocity of the
vertical stage, while continuously monitoring the lens position, the applied force and the contact radius. Experiments are performed at three different values of the
driving velocity $V=0.02,$ $0.002,$ $0.0002$ \textrm{mm/s}. Three contact
realizations have been carried out for each velocity.

\subsubsection{Experiments on rough samples}

PDMS\ samples were textured with a statistical distribution of spherical micro-asperities with the same radius of curvature. The
patterned surface was obtained by moulding PDMS\ in PolyMethylMethAcrylate
(PMMA) forms milled using ball-end mills with a radius of $100$ $\mathrm{\mu
m}$. In order to enhance adhesive effects, a smoothening of the spherical cavities of the PMMA\ molds has been
achieved by exposing them to a saturated \textrm{CHCl}$_{3}$ vapor for $30$
minutes. As detailed in Ref. \cite{ACITO2019}, such treatment leads to a slight increase in the radius of the spherical
bumps, up to a $10\%$ enhancement.

The patterned surface has been generated with a squared nominal area of $10$ 
$\mathrm{mm}^{2}$, where asperities are randomly distributed with a density
of $2\times 10^{7}$ $\mathrm{m}^{-2}$. The spherical caps present heights
distributed according to a Gaussian law with standard deviations $\sigma =5$ 
$\mathrm{\mu m}$.

Asperities are collocated with a non-overlapping constraint. For the considered surface density,
each contacting asperity behaves as an isolated spherical punch and
lateral interactions can be neglected as shown in Refs. \cite%
{ACITO2019,YASHIMA2015}. We stress that such assumption is no longer valid
when roughness on several length scales is considered like in the case of
self-affine fractal geometries \cite{ViolanoJKR,ViolanoDMT,PersAff}.

\begin{figure}[tbp]
\begin{center}
\includegraphics[width=15.0cm]{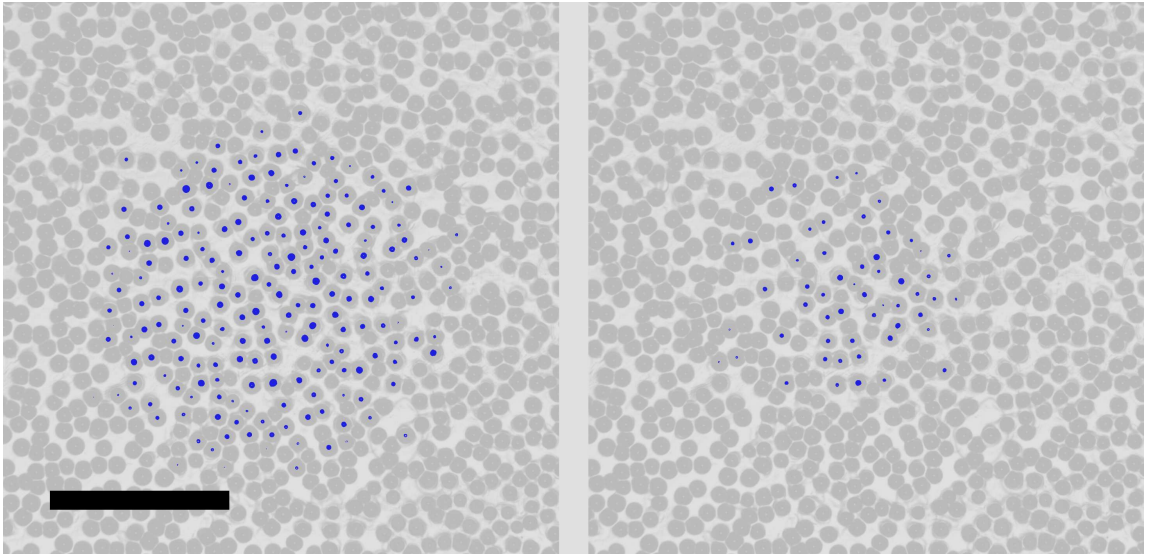}
\end{center}
\caption{Detachment of spherical micro-asperities. The contact spots (blue
disks) are detected after post-processing of the contact pictures. The length of the black rectangle is 1 mm.}
\label{Microspot}
\end{figure}

Fig. \ref{Microspot} shows images of contact micro-spots (blue disks) during
unloading. Accurate measurements of the area of contact spots is achieved by image processing after background removal. During unloading, experiments have been performed at
the same values of driving velocity $V$ used in the tests conducted on
smooth samples ($V=0.02,$ $0.002,$ $0.0002$ \textrm{mm/s}).

\section{Results}

\subsection{Smooth contact: macroscopic scale}

\begin{figure}[tbp]
\begin{center}
\includegraphics[width=15.0cm]{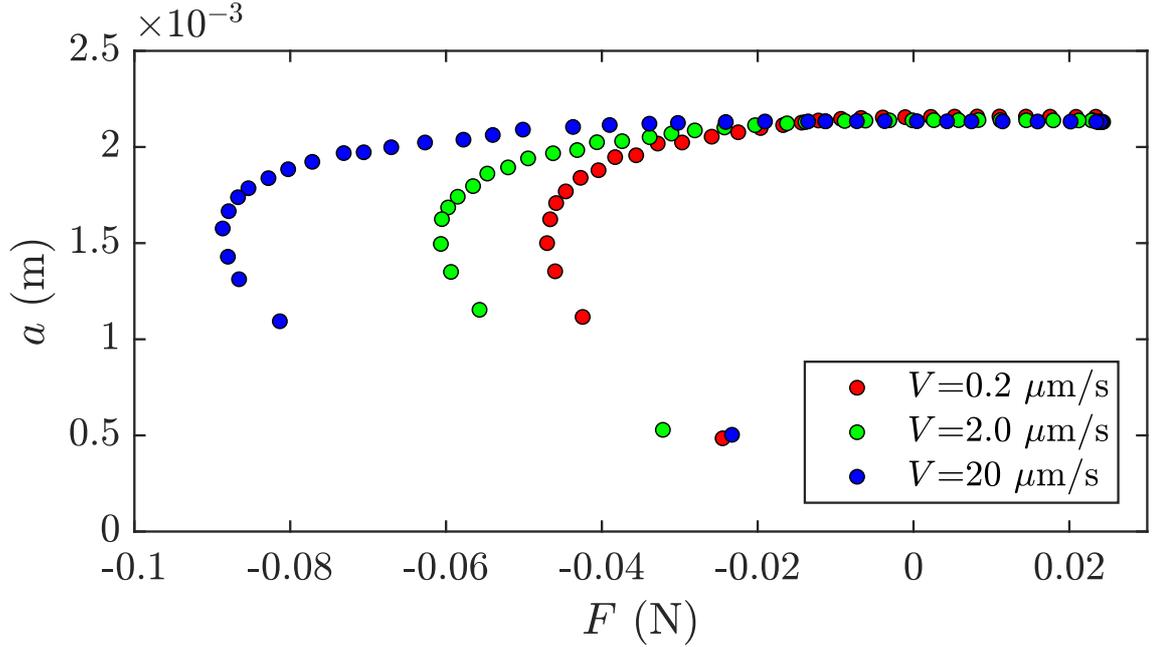}
\end{center}
\caption{The contact radius $a$ as a function of the applied load $F$.
Results are shown for different unloading velocities of the indenter $V=0.02$%
, $0.002$, $0.0002$ \textrm{mm/s}. For each velocity three tests have been
performed and the average values are reported in the plot.}
\label{Fmacro}
\end{figure}

Fig. \ref{Fmacro} shows the contact radius $a$ as a function of the applied
load $F$, during unloading. Results are obtained for different values of the
driving velocity $V$. Three contact tests were performed for each $V$ and
the average values are reported in the plot.

The detachment process is clearly rate-dependent as great adhesion
enhancement is observed by increasing $V$, as indicated by the increase in pull-off force.

\begin{figure}[tbp]
\begin{center}
\includegraphics[width=15.0cm]{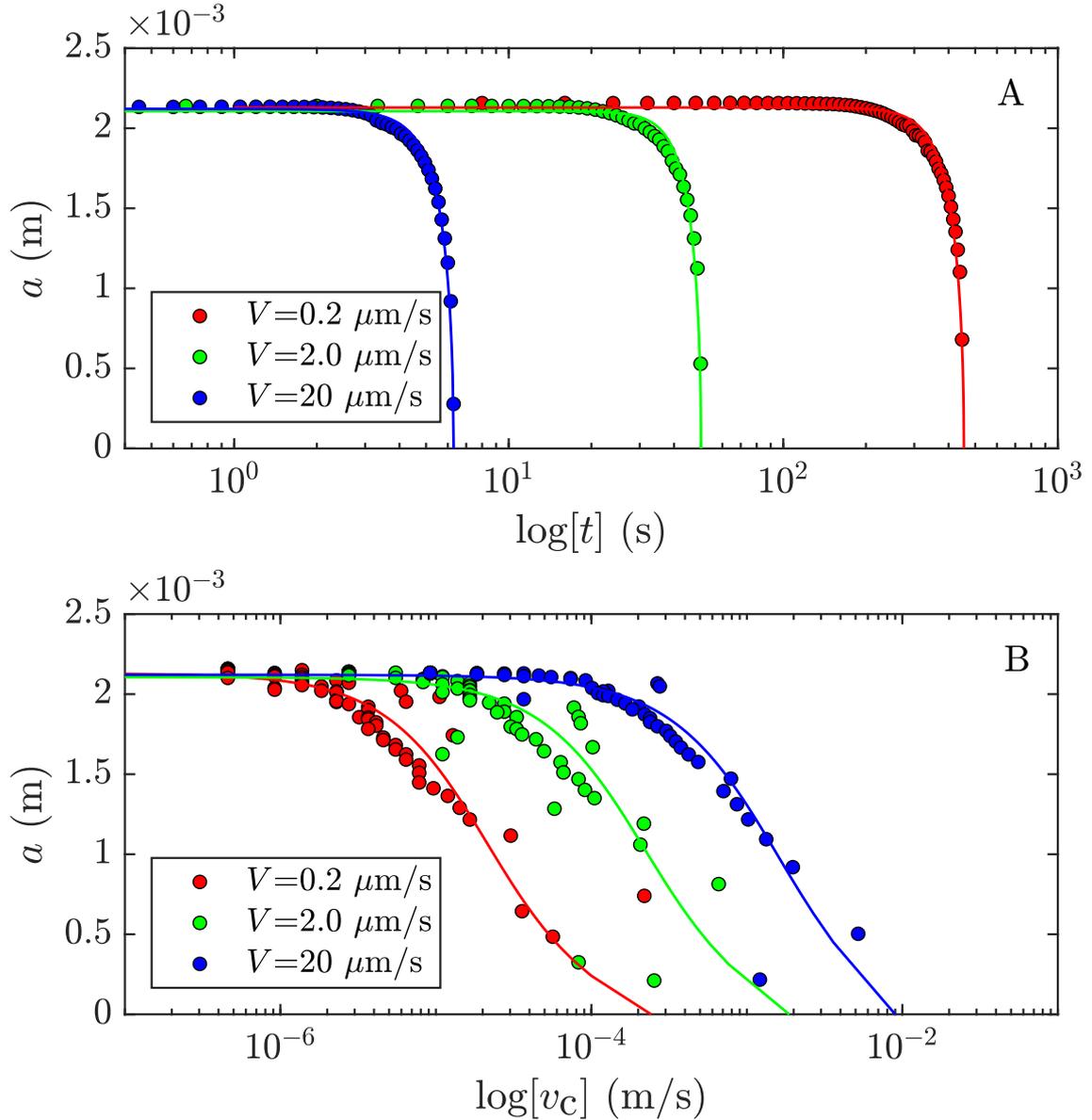}
\end{center}
\caption{A: The contact radius $a$ as a function of the time $t$. Results
show the average of three contact tests. Blue, green and red markers are
referred to the unloading velocities of the indenter $V=0.02$, $0.002$, $%
0.0002$ \textrm{mm/s}, respectively. Solid lines denote the fit of
experimental data. B: The contact radius $a$ as a function of the crack tip
velocity $v_{\mathrm{c}}$. Legend symbols are the same of A.}
\label{RVmacro}
\end{figure}

Fig. \ref{RVmacro}A shows, in a semilogarithmic representation, the
reduction of the contact radius $a$ with the time $t$. Experimental data are
fitted according to the following relation%
\begin{equation}
a(t)=p_{1}\sqrt{\left( 1-\frac{t}{t_{\mathrm{po}}+1}\right) ^{p_{2}}}
\label{at}
\end{equation}%
where $t_{\mathrm{po}}$ is the instant at which detachment occurs and $%
p_{1}$, $p_{2}$ are fitting parameters.

During detachment, the crack tip velocity can be easily obtained as $v_{%
\mathrm{c}}=-da/dt$. Fig. \ref{RVmacro}B shows the curves $a$ vs. $v_{%
\mathrm{c}}$ obtained in the experiments at different velocities. The
maximum value of $v_{\mathrm{c}}$ is reached when abrupt pull-off occurs.
Notice that when increasing $V$ of one order of magnitude the same
enhancement in $v_{\mathrm{c}}$ is also observed.

The effective work of adhesion $G$ is calculated as a function of measured contact line velocity by eq. (\ref{G3}) using the experimental value of the normal load $F$. As discussed by Barquins \cite{stiffness1983}, the resulting $G(v_{\mathrm{c}})$ relationship is unaffected by the machine compliance. The actual stiffness of the system is taken into account because the load $F$ is read from the deflection of the double cantilever beam. The same procedure has been recently used in Ref. \cite{baekPDMS}, where the effective work of adhesion is measured in spherical contact between a glass lens and PDMS blocks.

\begin{figure}[tbp]
\begin{center}
\includegraphics[width=15.0cm]{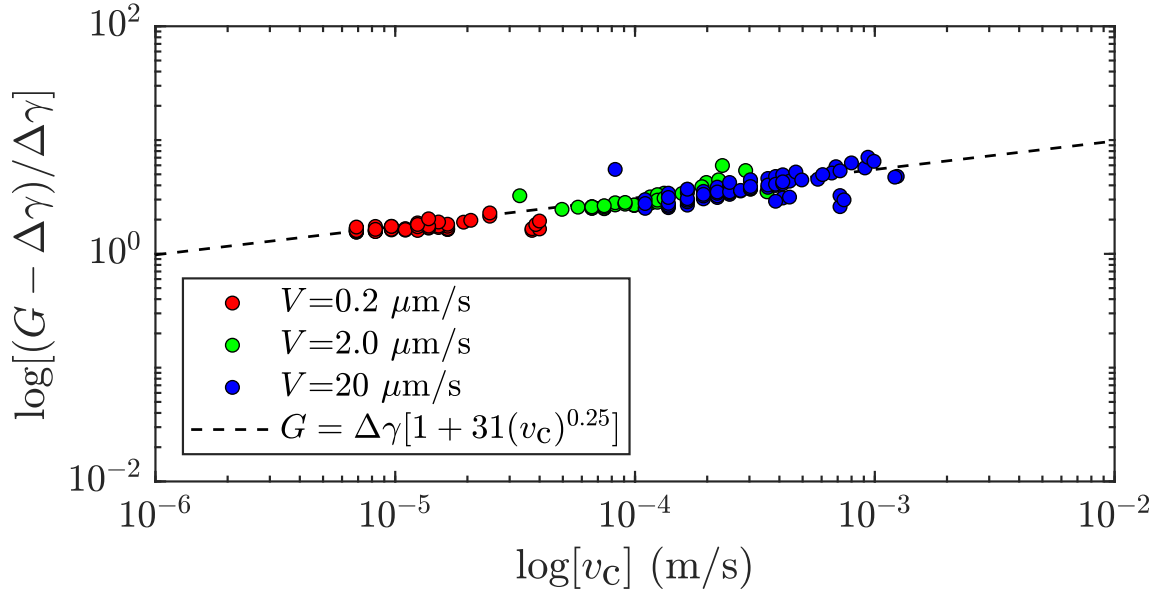}
\end{center}
\caption{The relative increase $\left( G-\Delta \protect\gamma _{0}\right)
/\Delta \protect\gamma _{0}$ as a function of the crack tip velocity $v_{%
\mathrm{c}}$. Results are shown for three contact realizations. Blue, green
and red markers are referred to the unloading velocities of the indenter $%
V=0.02$, $0.002$, $0.0002$ \textrm{mm/s}. The dotted line is the fit
obtained with eq. (\protect\ref{G2}).}
\label{Gmacro}
\end{figure}

Fig. \ref{Gmacro} shows the quantity $(G-\Delta \gamma _{\mathrm{0}})/\Delta \gamma _{\mathrm{0}}$ as a function of $v_{\mathrm{c}}$ in a double logarithmic
representation. Markers denote experimental data, while the dotted black
line is the fit obtained with eq. (\ref{G2}) using $c=31$ and $n=0.25$. Maugis $\&$ Barquins \cite{MB1978}
found $n=0.6$ for a viscoelastic polyurethane rubber. More recently, Lorenz et al.~\cite{Lorenz2013} found $n=0.19$ for polyurethane 
and $n=0.12$ for Sylgard 184 PDMS rubber. As discussed by Barthel and Fr\'etigny~\cite{Barthel2009}, the dependence of $G$ on the crack tip velocity can be related to the viscoelastic creep function of the solids. Accordingly, the fact that we found for the used Sylgard 184:Sylgard 527 mixture an exponent $n$ greater than for raw Sylgard 184 probably reflects the enhanced viscoelastic dissipation of the PDMS mixture.

The values of $c$ and $n$ can be used in eq. (\ref{beta}) to calculate the
values of $\beta $ required in Muller's model.
According to Muller's model, eq. (\ref{Meq}) can be numerically integrated to obtain $a({\delta})$. Assuming that $V$ is constant during unloading, the instant $t_{\mathrm{po}}$ at which pull-off occurs can be estimated by $t_{\mathrm{po}}=(\delta_{0}-\delta _{\mathrm{po}})/V$, where $\delta _{0}$ is the initial penetration and ${\delta}_{\textrm{po}}$ the jump-off distance. However, in our experiments we found that the actual unloading velocity $V_{\mathrm{act}}(t)$ is not constant as the spherical indenter is held to the stage using a compliant double cantilever beam. Due to the deflection of the beam, the velocity $V_{\mathrm{act}}$ of the lens can be different from the imposed velocity $V$. This is especially true for the experiments on smooth PDMS, where high forces can be achieved.
However this effect is quantified by the laser displacement sensor, which monitors the actual position $z(t)$ of the lens. Hence, the actual velocity is derived as $V_{\mathrm{act}}(t)=\Delta z/\Delta t$.
In the original Muller's model, the unloading velocity $V=-d\delta /dt$ is assumed to be constant.
However, we can modify eq. (\ref{beta}) by introducing the actual velocity $%
V_{\textrm{act}}(\delta )$ in the parameter $\beta$  
\begin{equation}
\beta =\left( \frac{6}{\pi }\right) ^{1/3}\left( \frac{4}{9}c\right) ^{1/n}V_{\textrm{act}}(%
\bar{\delta})\text{.}  \label{betanew}
\end{equation} 
where $V_{\textrm{act}}(\delta )$ is obtained by interpolating experimental data. 

The time required to move from $\delta _{0}$ to a generic $\delta$ is then calculated as%
\begin{equation}
t=\int_{\delta_{0}}^{\delta}\frac{\delta }{V_{\mathrm{act}}(\delta )}d\delta. 
\label{tempo}
\end{equation}

\begin{figure}[tbp]
\begin{center}
\includegraphics[width=15.0cm]{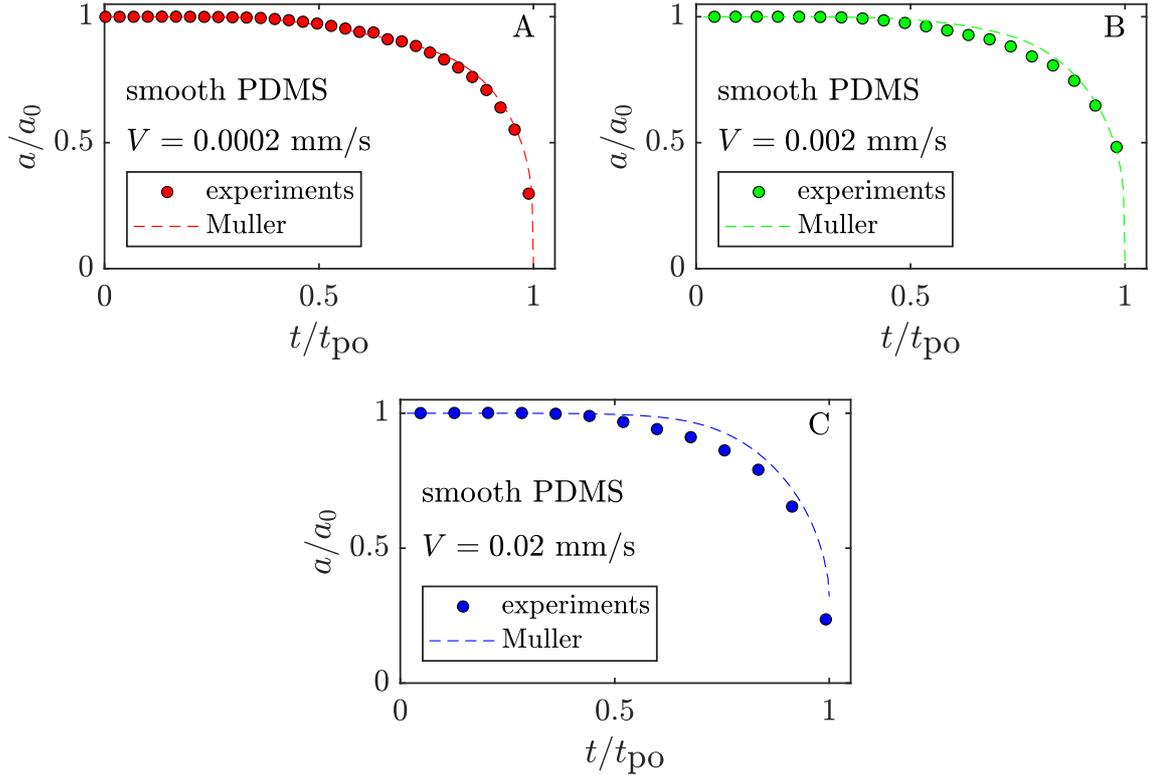}
\end{center}
\caption{A-C: The time dependence of the normalized contact radius $a(t)$. Results are shown for unloading velocity of the indenter $%
V=0.0002$, $0.002$, $0.02$ \textrm{mm/s} (figs. A,B,C respectively). Dashed lines denote Muller's model predictions, while markers experimental data, which are averaged on three contact realizations.}
\label{at_smooth}
\end{figure}

Results are shown in Figs. \ref{at_smooth}A-C, where the contact radius $a(t)$, normalized with respect to its initial value, is plotted against the time normalized with respect to the period required for pull-off to occur. Results are given for different unloading velocities and show a good agreement between experimental data and numerical predictions.

\subsection{Rough contact: microscopic scale}

\begin{figure}[tbp]
\begin{center}
\includegraphics[width=15.0cm]{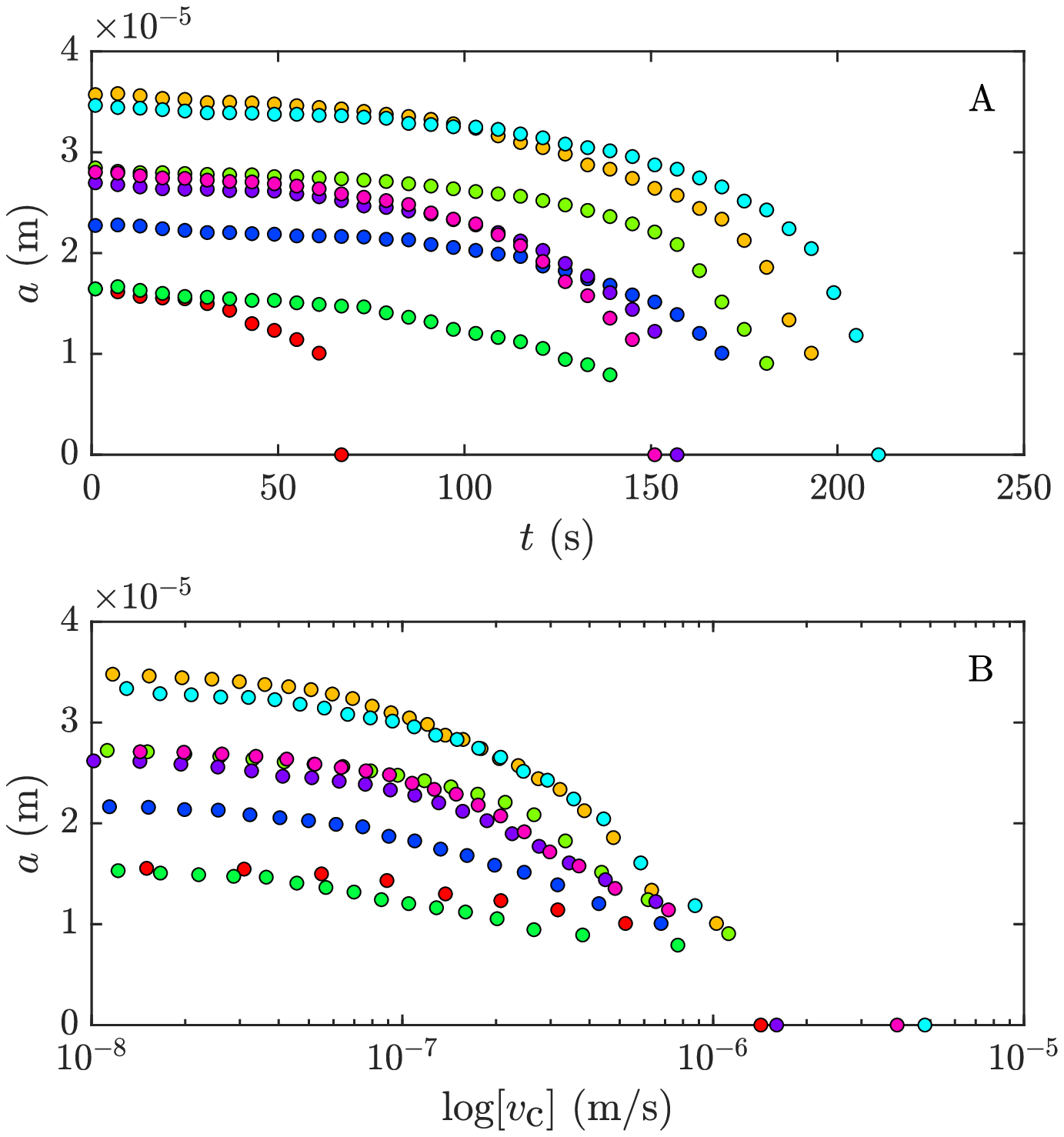}
\end{center}
\caption{A: The contact radius $a$ of micro-asperities as a function of time 
$t$. Markers denote experimental data for a selection of eight different micro-contacts. Unloading tests are
performed at $V=0.0002$ \textrm{mm/s}. B: The contact radius $a$ as a
function of the crack tip velocity $v_{\mathrm{c}}$.}
\label{V1micro}
\end{figure}

In the experiments performed on rough samples the number of micro-asperities
detected in contact at the end of the loading phase is around $160$.
However, for the sake of clarity, Figs. \ref{V1micro}A and \ref{V1micro}B
show the variation of the contact radius $a$ of $8$ micro-asperities in
terms of the time and the contact line velocity $v_{\mathrm{c}}$,
respectively. Results are shown for $V=2\times 10^{-4}$ \textrm{mm/s}. In
general, asperities with a larger value of the initial contact radius $a$
require a longer time to complete their detachment process.

Results on smooth and rough samples suggest the existence of scale
effects on both contact radius $a$ and detachment front velocity $v_{\mathrm{%
c}}$. For this reason, we rescale the above results introducing the factors $%
s_{\mathrm{a}}=a_{0\mathrm{macro}}/a_{0i}$, $s_{\mathrm{t}}=t_{\mathrm{%
po-macro}}/t_{\mathrm{po-}i}$ and $s_{\mathrm{v}}=s_{\mathrm{a}}/s_{\mathrm{t%
}}$. The quantity $a_{0\mathrm{macro}}$ is the initial value of the contact
radius measured at the macroscale on smooth PDMS samples at the end of the
loading phase (when unloading starts); $a_{0i}$ is instead the initial value
of the contact radius detected for the $i^{th}$ micro-asperity. Similarly, $%
t_{\mathrm{po-macro}}$ is the time at which pull-off occurs at the
macroscale (that is measured in the tests performed on smooth PDMS samples),
while $t_{\mathrm{po-}i}$ is the time required (and measured in the tests on
rough PDMS samples) to detach the $i^{th}$ micro-asperity.

Therefore, contact radius $a$, crack tip velocity $v_{\mathrm{c}}$ and time
are rescaled with the above factors. The new curves are given in Fig. \ref%
{RVmicro} for three different driving velocities $V$ and in a semi-log plot.
Solid lines denote the curves obtained at the macroscopic scale in the
experiments conducted on smooth substrate. Dashed lines identify the curves
measured for each micro-asperity during the detachment tests performed on
the rough PDMS samples. All curves obtained on the contact microspots almost
collapse on the curves measured at the macroscale (smooth samples). Such
result suggests that the distributions of the actual crack-tip velocities $%
v_{\mathrm{c}}$, which are achieved locally at micro-contact, scale during
contact unloading. This, in turn, suggests that  the parameters of Muller's model identified at the macroscale can be applied to the microcontacts.

\begin{figure}[tbp]
\begin{center}
\includegraphics[width=15.cm]{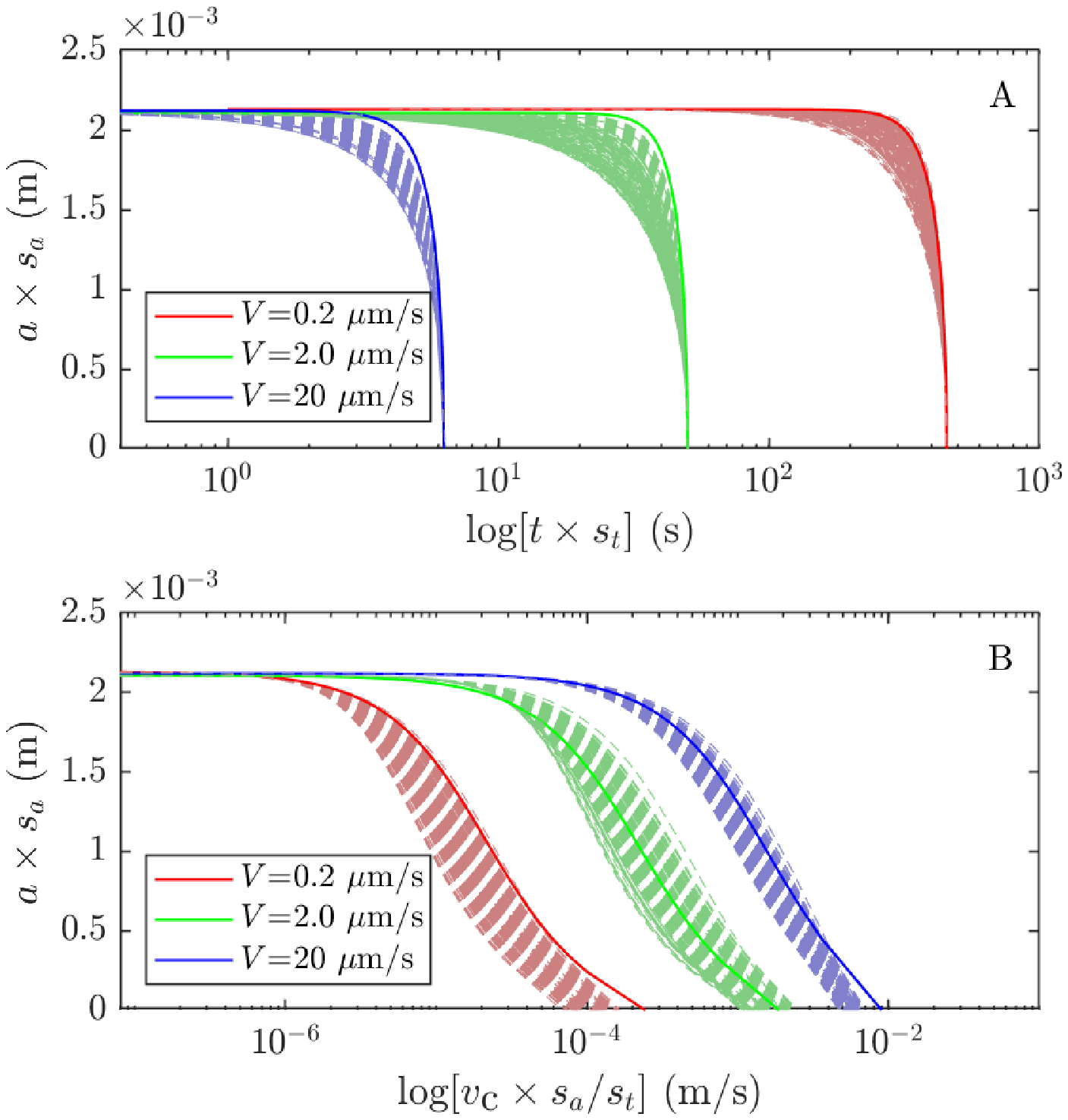}
\end{center}
\caption{A: The contact radius $a\times s_{a}$ \ as a function of time $%
t\times s_{t}$ (semilog scale). Solid lines denote the smooth macro-spot
detachment curve. Dashed lines denote the detachment curves of $160$
micro-asperities. Red, Green and Blue curves refer to $V=0.2,$ $2.0,$ $20,$ $%
\mathrm{\protect\mu m/s}$. A: B: The contact radius $a\times s_{a}$ \ as a
function of the crack tip velocities $v_{\mathrm{c}}\times s_{a}/s_{t}$ (semilog
scale). Legend symbols are the same of A.}
\label{RVmicro}
\end{figure}

Such an assumption finds also its motivation in recent results by Lorentz et
al. \cite{Lorenz2013}, who performed adhesion experiments on smooth spheres
of different radii (ranging from $R\approx 3$ \textrm{mm} to $R=46.5$ 
\textrm{mm}) and different materials. They deduced $\Delta \gamma _{\mathrm{%
eff}}$ as a function of the contact line velocity $v_{\mathrm{c}}$ using the
JKR theory and observed that the experimental data exhibited the same
velocity dependence as calculated by eq. (\ref{G1}) for $v_{\mathrm{c}%
}<10^{-4}$ \textrm{m/s} (which corresponds to the range of crack tip
velocities measured in our experiments on micro-spots).

\begin{figure}[tbp]
\begin{center}
\includegraphics[width=15.0cm]{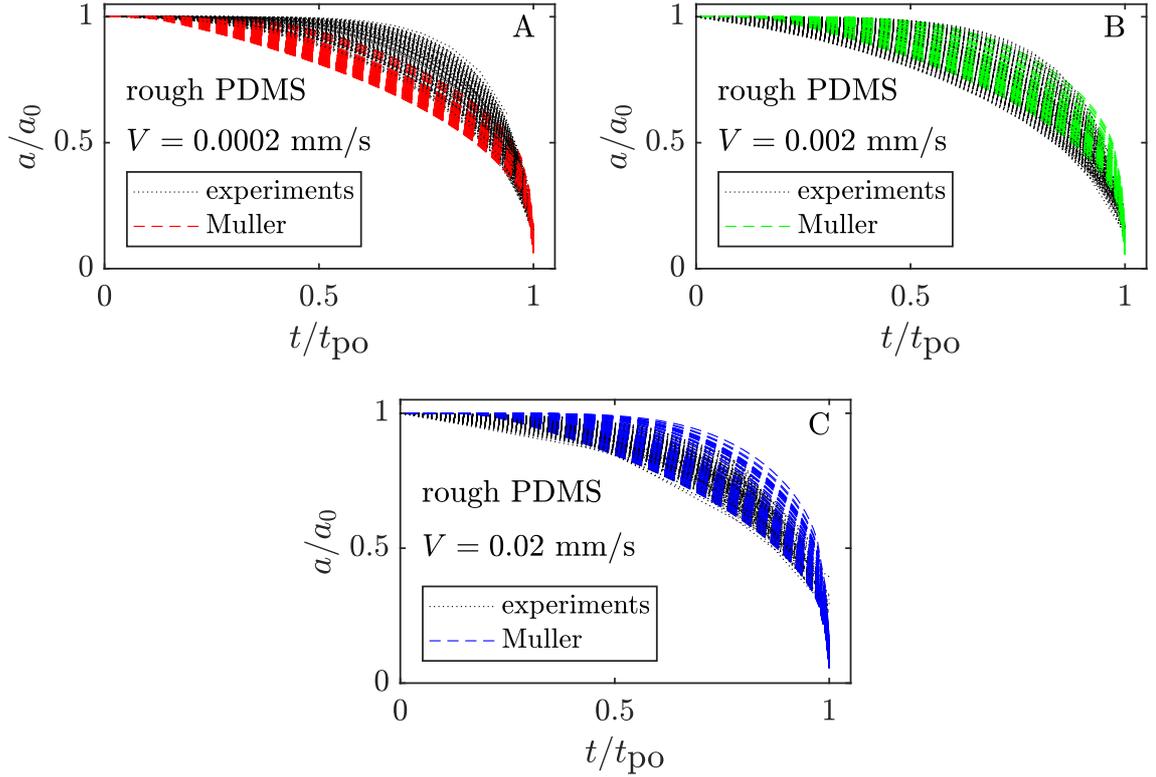}
\end{center}
\caption{A-C: The time dependence of the normalized contact radius $a(t)$ for micro-contacts. Results are shown for unloading velocity of the indenter $V=0.0002$, $0.002$, $0.02$ \textrm{mm/s} (figs. A,B,C respectively). Black dotted lines denote experimental data, while colored dashed lines the Muller's model predictions.}
\label{at_rough}
\end{figure}

\begin{figure}[tbp]
\begin{center}
\includegraphics[width=15.0cm]{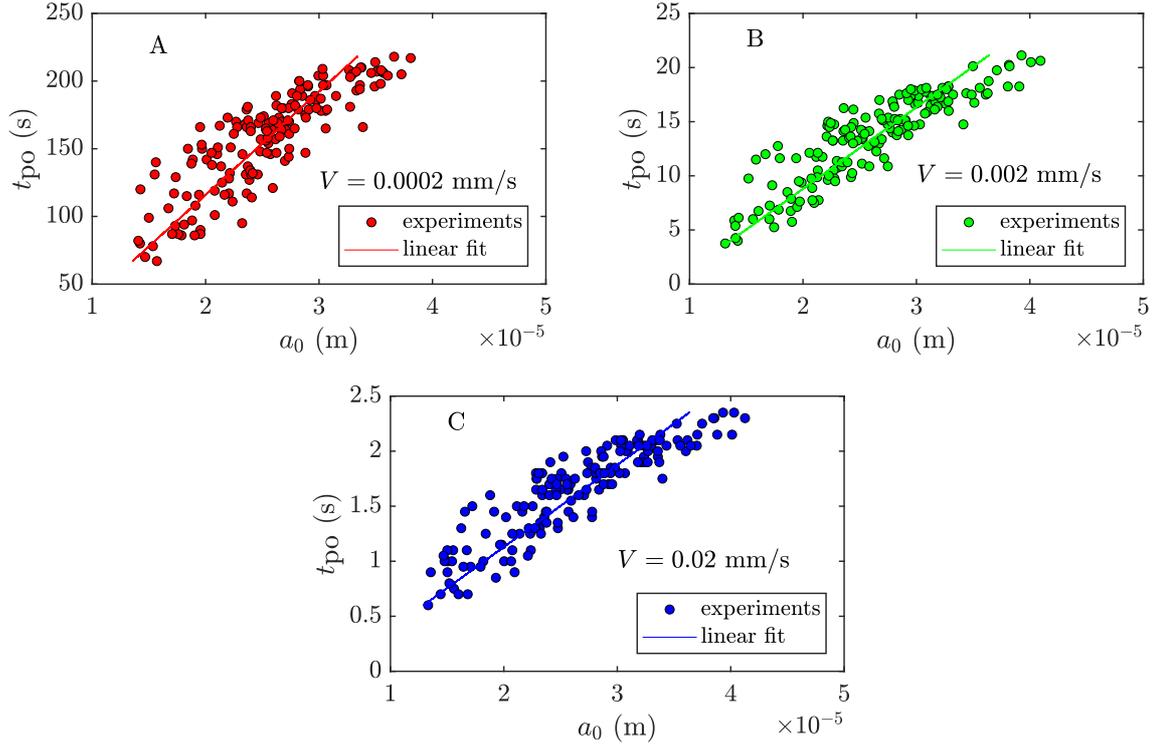}
\end{center}
\caption{A-C: The pull-off time $t_{\mathrm{po}}$ as a function of the initial contact radius $a_{0}$. Results are shown for rough PDMS and unloading velocity of the indenter $V=0.0002$, $0.002$, $0.02$ \textrm{mm/s} (figs. A,B,C respectively). Markers denote experimental data, while colored lines the corresponding linear fit; the $R^{2}$ value is 0.73, 0.77 and 0.78 (figs. A,B,C respectively).}
\label{at_trend}
\end{figure}

The same plots given in Figs. \ref{at_smooth} are reported in Figs. \ref{at_rough} for each of the micro-contacts detected during the unloading phase. A satisfactory good agreement is found between experimental data and numerical predictions, which are obtained with the "macroscale" values of $c$ and $n$. Also in this case, in the Muller's model, we have introduced the actual value of the unloading velocity, which is however constant and slightly lower than the imposed one ($V_{\textrm{act}}=0.8V$).
In the $a-t$ curves, we can distinguish a period of time where stick adhesion is observed with an almost constant contact radius (i.e. the contact line velocity is zero). This "stick time" is negligibly influenced by the initial value $a_{0}$ of the contact radius \cite{Barthel2003}. For this reason, when increasing $a_{0}$, the $a-t$ curve does not scale homothetically and a master curve cannot be found. This explains the scatter in fig. \ref{at_rough}. However, an increasing trend of the pull-off time with $a_{0}$ can be observed in our experimental data, as shown in figs. \ref{at_trend}A-C, where results are presented for different unloading velocities. Anyway, as data are strongly scattered, a clear law of this increasing trend is not identified.

\section*{Conclusions}

In this paper, we have investigated the pull-off behavior of a rough contact interface between a smooth glass lens and a nominally flat viscoelastic substrate patterned with a height distribution of spherical micro-asperities. In the absence of any elastic coupling between micro-contacts, this system allows to measure simultaneously the pull-off behavior of a collection of micro-asperities contacts differing in their initial (equilibrium) adhesive contact radius. From a comparison with macroscale pull-off experiments, it also offers the possibility to investigate the occurrence of scale effects in dissipative processes involved in adhesion.\\
Results show that the contact radius almost scales according to the ratio $%
s_{\mathrm{a}}=a_{0\mathrm{macro}}/a_{0\mathrm{micro}}$, being $a_{0}$ the initial radius measured at the beginning of the unloading process. Similarly, the contact line velocity $v_{\mathrm{c}}$\ is found scaling with a factor $s_{%
\mathrm{v}}$ depending on the ratio $s_{\mathrm{a}}/s_{\mathrm{t}}$, where $%
s_{\mathrm{t}}=t_{\mathrm{po-macro}}/t_{\mathrm{po-micro}}$ and $t_{\mathrm{%
po}}$ is the time required for the pull-off to take place.
Such results suggest that the nature of the dissipative processes involved in the pull-off of the adhesive contacts is almost scale independent from the millimeter size down to a few tens of micrometers. In other words, the assumption that viscoelastic losses are localized near the contact line in a region small with respect to the contact size remains valid at the micro-scale. Moving from this consideration, a simple theoretical procedure can be derived to evaluate the evolution of the contact radius $a$ of micro-asperities contacts during unloading.

\end{document}